\documentclass[preprint,12pt]{elsarticle}

\usepackage{amssymb}

\usepackage{subcaption}
\usepackage{hyperref}
\usepackage{cleveref}

\usepackage{tabularx}
\usepackage[draft]{changes} 
\usepackage{lineno}
\definechangesauthor[name={CD}, color=orange]{CD}

\journal{Nuclear Inst. and Methods in Physics Research, A}

\begin{document}

\begin{frontmatter}



\title{Background in Low Earth Orbiting Cherenkov Detectors, and Mitigation Strategies.}


\author[a]{C. S. W. Davis}
\author[a]{F. Lei\corref{cor1}}
\author[a]{K. Ryden}
\author[a,b]{C. Dyer}
\author[c]{G. Santin}
\author[c]{P. Jiggens}
\author[d]{M. Heil}

\cortext[cor1]{Corresponding author: f.lei@surrey.ac.uk}

\affiliation[a]{organization={Space Environment and Protection Group, Surrey Space Center, University of Surrey},
            addressline={Stag Hill, University Campus}, 
            city={Guildford},
            postcode={GU2 7XH}, 
            state={Surrey},
            country={United Kingdom}}

\affiliation[b]{organization={CSDRadConsultancy},
addressline={Fleet, GU52 6PZ, United Kingdom}}

\affiliation[c]{organization={European Space Research and Technology Centre},
            addressline={ European Space Agency, Noordwijk, Netherlands}}

\affiliation[d]{organization={European Space Operations Centre},
            addressline={European Space Agency, Darmstadt, Germany}}

\begin{abstract}
Cherenkov detectors have been used in space missions for many decades, and for a variety of purposes, including for example, for Galactic Cosmic Ray (GCR) and Solar Energetic Particle (SEP) measurements. Cherenkov detectors are sensitive to many types of particles that are present in the environment of space, including gamma rays, trapped particles and cosmic particles, and each particle component acts as essentially a background when trying to view another specific particle component. In this research, GRAS/Geant4 simulations were performed to characterise the count rates that a simple Cherenkov detector design would experience in a low Earth orbit, and we find that Cherenkov count rates due to most particle components vary significantly depending on many different factors, including the location in the orbit, the date of the orbit, whether or not the detector is within the van Allen belts, and whether or not a solar particle event is occurring. We find that a small Cherenkov detector is readily able to gather detailed data on both trapped particles and spectral information during Ground-Level Enhancements. We also investigate the use of coincidence as a method to remove count rates due to trapped particles and delta electrons, finding that this method is generally very effective for resolving count rates due to GLEs amongst intense trapped particle environments, but that some Cherenkov count rates due to trapped particles are still observed in the simulated south Atlantic anomaly region. 
\end{abstract}


\begin{highlights}
\item GRAS simulations of a 1cm$\times$1cm$\times$1cm fused silica Cherenkov radiator with SiPM detector find that the instrument could readily measure solar energetic particles, cosmic rays and trapped electrons and protons in low Earth orbit
\item The use of different photon channels and coincidence modes of operation can reduce trapped particle count rates, and allow for investigations of spectral properties of incoming particles
\item Count rates during the South Atlantic Anomaly due to high energy protons at energies below the Cherenkov threshold were found to be significant, due to delta electrons
\end{highlights}


\end{frontmatter}


\section{Introduction}
\label{sec:Introduction}

Cherenkov detectors utilise the fundamental physics of Cherenkov radiation \cite{jackson1999classical, vcerenkov1937visible}, which occurs when a charged particle traveling through a material is faster than the speed of light in that material. As Cherenkov light production can therefore only be produced by particles at relativistic speeds, they intrinsically don't experience count rates due to particles with low kinetic energies. Cherenkov detectors therefore exclude many types of low energy particle background, making them a useful type of detector for scientific observations involving high-energy charged particles, and are commonly associated with high-energy particle physics measurements. 

The feature of Cherenkov detectors to exclude particle counts below specific threshold kinetic energies is particularly useful in the environment of space, where there is typically a large quantity of charged low-energy particles, and the particle kinetic energy spectra for particular particle species are typically low-energy dominated shapes such as power laws. These low-energy components can easily obscure other particle components, if they aren't removed.

While Cherenkov detectors exclude particle backgrounds due to low energy charged particles, there are still many particle components that a space-based Cherenkov detector could and will experience. In any scientific observation aiming to view one of these components, all other particle components will act as background, and reduce the observation quality. This means that understanding the count rates the detector will receive from the full range of different particle components during the detector and mission design is important, so that the observation quality can be assessed, and background mitigation strategies can be designed. 

However, to our knowledge, there have not been many studies published assessing the full background that would be present in a generic Cherenkov detector in Earth orbits through space environment simulations, and background mitigation strategies. Although not specific to Cherenkov detectors, some specific examples of excellent review and simulation work on the background in specifically space-based gamma ray observatories include work by Campana et al.\cite{campana2013background}, Cumani et al. \cite{cumani2019background}, Arneodo et al.\cite{arneodo2021review}, Galgóczi et al. \cite{galgoczi2021simulations}.

The research presented in this paper characterises the particle components that are likely to be detected by a simple space-based Cherenkov detector in a low-Earth orbit through GRAS\cite{santin2005gras} / Geant4\cite{agostinelli2003geant4} simulations as a function of location. We focus on the context of a detector attempting to observe solar energetic particle spectra, partly because this research with performed as part of ESA's HEPI project, however the results of this work can be applied in general to space-based Cherenkov detectors. The effectiveness of several methods such as the use of coincidence and pulse height distribution shape are also investigated as ways to separate the particle components.


To understand the effect that a particular particle component has on detector count rate, it is useful to consider the context of the well-known Frank-Tamm formula\cite{frank1937coherent}. The number of photons produced by a particular particle species through Cherenkov radiation per unit distance, per unit wavelength, can be calculated in terms of incoming particle kinetic energy to be

\begin{equation}
\label{eq:CherenkovEquation}
    \frac{\partial^2 N}{\partial x \partial \lambda} = \frac{2\pi \alpha}{\lambda^2} \left(1-\frac{(E_k + m_0 c^2)^2}{E_k n^2 (E_k + 2 m_0 c^2)}\right)
\end{equation}

Where $E_k$ is the particle kinetic energy, $m_0$ is the particle rest mass, $c$ is the speed of light, $n$ is the material refractive index, $\alpha=\frac{1}{137}$ is the fine structure constant, $x$ is distance through the material, and $\lambda$ is photon wavelength. The minimum kinetic energy required to breach the speed of light in the material (equivalent to setting \cref{eq:CherenkovEquation} equal to 0) can be calculated to be

\begin{equation}
    E_{k,min} = m_0 c^2 (\frac{n}{\sqrt{n^2-1}} - 1)
\end{equation}

which gives, for example, minimum kinetic energies of $E_{k,min}$ = 402 MeV for protons and $E_{k,min}$ = 219 keV for electrons, for a refractive index of $n$ = 1.4. This means that through the choosing of refractive indices, a space-based Cherenkov detector can be made sensitive to exclusively particles with kinetic energies greater than a certain threshold. A good plot of this is shown by Mazur et al.\cite{mazur2014relativistic}, who show the response of the Relativistic Proton Spectrometer (which used Cherenkov detectors) on the NASA Van Allen Probes spacecraft to protons with different energies. This makes Cherenkov detectors ideal for the measurement of protons with energies above several hundred MeV in energy (which corresponds to the energies which happen to be most impactful in terms of radiation dose rates on Earth). However Cherenkov detectors are still susceptible to several potential background sources. 

Background is present in almost all scientific measurements, and can cause both statistical and systematic uncertainties. Statistical uncertainties are caused specifically by non-signal count rates that can be calculated and removed from observations, but only in aggregate. This typically acts to reduce the quality of scientific measurements through reducing signal-to-noise ratio, which leads to longer required observation times, that can sometimes become unfeasibly large in the case of large background count rates. In contrast, systematic uncertainties arise from sources of background where it is not possible to accurately calculate or remove the non-signal count rates from observations, which can lead to observational data being skewed or even negated completely. Highly variable background sources tend to fall into this category.

There are numerous sources of count rates that could be observed by a space-based Cherenkov detector in a low-Earth orbit, each of which could be considered either as an interesting scientific signal, or as a form of background for observations of other count rate sources. Some of the most important \cite{cumani2019background} include:
\begin{itemize}
    \item Galactic Cosmic Rays (GCRs): galactic cosmic protons \cite{norbury2019advances,Schimmerling2011,matthia2013ready,davis2021simulation} originate from outside of the solar system, and are approximately isotropic outside of magnetospheres. Galactic cosmic proton spectra vary with solar cycle but do not typically vary significantly day-to-day (although minor fluctuations do occur due to a range of complex physical phenomena) unless a Forbush decrease \cite{forbush1937effects} is occurring, although they do vary across Earth-centered latitudes and longitudes due to Earth's geomagnetic shielding (which is also variable). Cosmic ions of other particle species will also cause count rates, but the spectra for these decreases rapidly with increasing atomic number \cite{perinati2012radiation} (as predicted by models such as CREME96 \cite{tylka1997creme96} for instance). Note that while cosmic electrons are often considered too low in intensity to be significant for spacecraft, compared to other particle species such as cosmic rays, it is possible that cosmic electrons could consistent a significant signal for space-based Cherenkov detectors in regions where Earth's magnetosphere doesn't provide shielding. This will be discussed further in the results section of this paper.
    \item Trapped protons: protons trapped in Earth's magnetic field, primarily in Earth's inner van Allen belt  \cite{norbury2019advances,Schimmerling2011,sawyer1976ap8}. Spacecraft travelling through the South Atlantic Anomaly (SAA) will therefore experience significant fluxes of these. Both these and galactic cosmic protons can reach energies of several hundred MeV and above, making them challenging to shield with passive, material-based shielding.
    \item Trapped electrons: electrons trapped in Earth's magnetic field, these are present in both the inner and outer van Allen belt \cite{norbury2019advances,Schimmerling2011,vette1991ae8}. These are challenging to model, as they may vary significantly on short timescales. These are therefore present in both the South Atlantic Anomaly (SAA), and in the polar `horns' regions at high and low Earth latitudes at low Earth orbits.
    \item Delta electrons: electrons generated by protons from other sources. Also sometimes referred to as 'secondary' electrons or 'knock-on' electrons, depending on the context \cite{sternglass1957theory,davis2021simulation}. As protons pass through both shielding and Cherenkov radiators, they can produce electrons through interactions with materials, which then could be high enough in kinetic energy to breach the electron Cherenkov threshold, even if the original proton wasn't high energy enough to breach the proton threshold. Technically these secondary particles are registered as part of count rates due to other primary particle sources, and resolving count rates due to secondary particles separately from the primaries that generate them is challenging.
    \item Cosmic gamma rays: Cherenkov detectors are frequently used in gamma ray observatories to detect gamma ray bursts \cite{murphy2021compact,cumani2019background}. The interaction process for this is that gamma rays can Compton scatter in Cherenkov radiators, producing electrons that can breach the Cherenkov threshold. Even though gamma ray bursts are relatively short-lived, there is a cosmic gamma ray background that is ever-present, and which could act as a background to other observations.
    \item Albedo particles: These are particles produced in Earth's atmosphere by atmospheric interactions with cosmic ions \cite{murphy2021compact,cumani2019background}. These therefore vary with cosmic ion spectra, which as discussed previously varies with Earth's geomagnetic shielding at a particular location.
\end{itemize}

 The research described in the rest of this paper discusses simulation work that was performed to assess the signals/backgrounds that could be present in a simple cubic fused silica Cherenkov radiator, coupled to a silicon photomultiplier. The mitigation strategies to remove background species that are tested are the use of a separate coincidence detector to remove non-penetrating particle count rates, and the use of photon thresholds/channels as a way to preferentially select events from non-electron sources. This research was performed as part of the HEPI project\cite{ONeill2025HEPI}, and will therefore focus on the detection of cosmic and solar protons, with other sources as background. However the general results presented here will likely be of interest for the scientific context of measuring other particle species too, such as gamma ray burst detection and trapped particle observations.

\section{The Specific Application of Cherenkov Detectors to Solar Energetic Particle Detection}
\label{sec:SEP_detection_intro}

Solar Energetic Protons (SEPs) and galactic cosmic protons with energies above several hundred MeV in energy can cause significant issues in both Earth-based and space-based systems \cite{shea2012space}. For instance, in space systems these protons can cause both long term damage to electronics and instantaneous impacts due to single event effects and ionisation. The effects of protons at these energies are challenging to mitigate, as the shielding required to block galactic cosmic protons at these energies is unfeasibly large (3.7~cm of aluminium is required just to block protons with 100~MeV of kinetic energy \cite{shen2018nist}). The Earth-based effects of solar and galactic cosmic protons include heightened radiation exposure for pilots and crew in aircraft, as well as aircraft electronics \cite{dyer1999cosmic, dyer2002radiation, hubert2020impact, prado2013effects, larsen2025study}. While the spectra of galactic cosmic protons at GeV energies can be reasonably evaluated using numerous modern models \cite{norbury2018comparison, norman2016evaluating,matthia2013ready,davis_cosraymodifiediso_nodate}, at least during non-Forbrush decrease time periods, the spectra and angular distribution of Ground-Level Enhancements (which are induced by large solar energetic proton events) remains extremely challenging to model. Some models for calculating dose rates during solar particle events include the isotropic MAIRE+\cite{hands2022new}, AVIDOS\cite{latocha2009avidos}, CARI-7\cite{copeland2017cari} and NAIRAS\cite{mertens2013nairas} models, and anisotropic models such as WASAVIES\cite{sato2018real}, CRAC:DOMO\cite{mishev2015computation}, SIGLE\cite{lantos2004semi} and AniMAIRE\cite{davis2024animaire}, however dose rates produced by all models can vary by orders of magnitude. This is primarily because Ground-Level Enhancements (GLEs) only occur once every couple of years, and means that there is intrinsically limited data available to study these events. The complexity of both the Sun, heliosphere, and Earth's magnetic field means that it is also challenging to explore these events through modeling too.

The relative scarcity of GLE solar energetic particle events means that maximising the data available from every event is important. Currently the most frequently used system for measurements of GLEs and cosmic particle fluxes is the global neutron monitor network \cite{mishev2020current}. This is a network of about 50 neutron monitors positioned around the Earth at specific locations such that they experience a wide variety of geomagnetic cut-off rigidities. A lot of research has been performed over many decades to relate neutron monitor count rates to space-based solar and galactic cosmic proton spectra \cite{koldobskiy2022fluences,larsen2023analysis,hands2022new,latocha2009avidos,baird2023potential}. However, fundamentally standard neutron monitors are only able to provide a single integral count rate proxy at one location on Earth's surface (albeit reliably, with high statistics and with direct relations to the radiation effects in Earth's atmosphere, among other advantages). In contrast, space-based measurements provide more direct in-situ measurements of spectra, and are ideal for validating these measurements and allowing for quantities like spectra and pitch angle distributions to be more directly measured, potentially with a single detector and with arguably less reliance on modeling.

There have been a number of space missions over the last few decades which can be used concurrently with ground-based neutron monitor data. Some well known detectors which can measure protons of several hundred MeV to GeV include AMS-02\cite{aguilar2021alpha}, PAMELA \cite{menn2013pamela} and many of the instruments that have been developed for the GOES series of satellites over the decades (e.g. the EPS instruments, EPEAD instruments, HEPAD instruments, and SGPS instruments \cite{sellers1996design, kress2020goes, kress2021observations, bruno2017calibration, rodriguez2023goes}). 

The GOES HEPAD series \cite{sellers1996design} were in fact Cherenkov detectors, with radiators made of quartz, and with additional front and back solid state detectors. The HEPADs were able to resolve protons with kinetic energies $>$970~MeV in the highest of its four channels. GOES satellites are positioned at geostationary orbit, and therefore should be expected to be exposed to nearly the full spectra of radiation-inducing cosmic and solar particles\cite{rodriguez2023goes}. The HEPADs were designed as telescopes; to provide directional discrimination of incoming particles, however contamination from out-of-acceptance particles proved an issue for this \cite{bruno2017calibration}.

Other Cherenkov detectors that have been used, and are being planned, for the space-based context of detecting cosmic and solar protons include detectors such as ARIEL-1 \cite{durney1964energy}, POGO \cite{Bingham1966}, HEOS-1 \cite{dyer1973possible, dyer1973studies}, OV1-20 \cite{nasaNASANSSDCA}, MEH/ISEE-3 \cite{moses1987jovian}, CRNC \cite{garcia1977age}, C2 \cite{bouffard1982heao}, KET \cite{simpson1992ulysses}, HET \cite{rodriguez2020energetic}, CHARMS \cite{2022cosp...44.3548S}. HEOS-1 was able to detect cosmic ray anisotropy for kinetic energies $>$360~MeV, and was able to detect numerous solar particle events of 1969 at high energies. Connell et al. have also proposed a novel Cherenkov detector utilising diamond as a radiator \cite{connell2019novel}.

A recent Cherenkov detector relevant to the work in this paper is the Relativistic Proton Spectrometer (RPS) \cite{mazur2023relativistic} aboard the Van Allen Probe spacecraft. The RPS consisted of 12 solid state detectors combined with a Cherenkov radiator. The RPS passed through the van Allen belts during its orbit, in an elliptical orbit about Earth, and picked up high count rates in both radiation belts. The detectors were able to detect trapped electrons, and trapped and galactic cosmic protons \cite{looper2021relativistic}, and were able to separate distributions of them algorithmically. Interesting differences were found between experimental data and simulations, possibly implying for instance the presence of an unaccounted for group of leptons in the data. RPS-B was also able to observe the SEP event of September 2017 (which included GLE72) \cite{o2018solar}. Reconstructions of spectra were made using this data, and it was found that at least in the case of this event, cut-off models were not able to account for the observed cut-off rigidities found in the data. The relatively complex structure of RPS and specifics of the algorithm RPS used to mitigate some forms of background means the background it experiences will likely be different to other bespoke designs, but nevertheless it is a useful and instructive example of a Cherenkov detector used to measure the space-based particle composition.

Each of these space-based Cherenkov detectors described here had very different designs in both radiator material, geometry and signal processing algorithms, much of which were part of bespoke and large-scale individual missions, meaning that every one of these detectors would experience different particle signals. 

The results described in the rest of this paper focus on a simple cubic fused silica radiator. This is a simpler design than many of the above described space-based Cherenkov missions, many of which used multiple radiators in variable arrangements and structures. However the relative simplicity of the simulated geometry used in this paper allows for easier qualitative understanding of results and provided the advantage of being easier and quicker to simulate. It is possible to qualitatively extend the results in this paper to other geometries, and for cubesat or constellation style missions, such as the HEPI project that this work was performed for. The simulations in this paper actually formed part of the basis for work recently published in O'Neill et al. (2025a), O'Neill et al. (2025b) \cite{o2025characterisation, ONeill2025HEPI}, where a very similar design to the design simulated in this paper was tested at the TRIUMF Proton Irradiation Facility (PIF). Photos of the setup as well as real-world device performance metrics can be found in the publication, as well as more information about the project.

\section{Simulation Setup}
\label{sec:Simulation_Setup}

The Geant4 Radiation Analysis for Space (GRAS) software\cite{santin2005gras} - version 6.00 beta with Geant4 version 10.7p04\cite{agostinelli2003geant4} - was used to perform simulations. To run simulations, a Python wrapper for GRAS was developed that was capable of running the Cherenkov simulations for any arbitrary spectrum or a list of spectra and for a range of different simulation geometry structures. This Python wrapper was used for all the simulations and analysis in this paper, and can be found at https://github.com/ssc-maire/SpaceCherenkovSimulator \cite{SpaceCherenkovSimulator}.

\begin{figure}
    \centering
    \includegraphics[width=\textwidth]{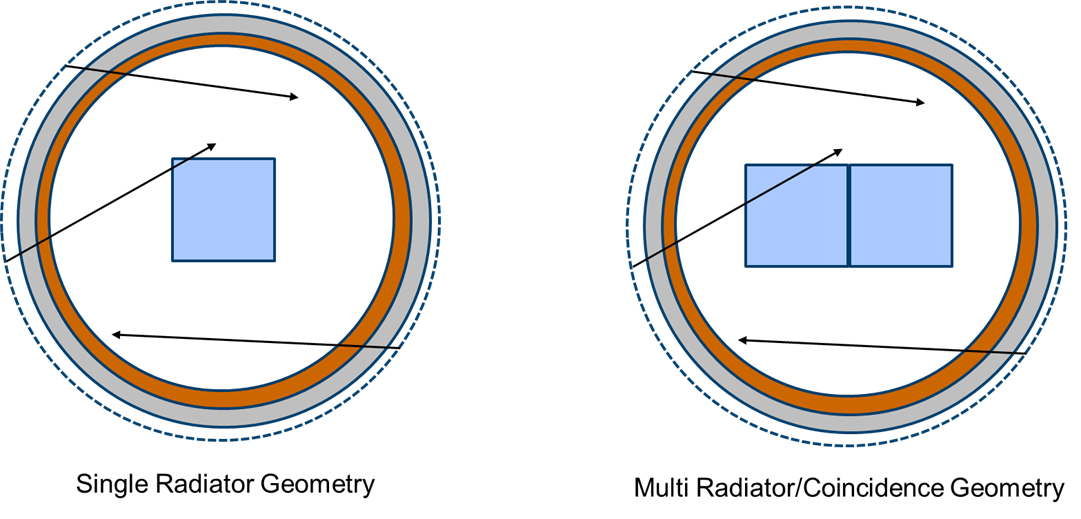}
    \caption{The two geometries that were simulated in GRAS in this work. In both cases, the shielding used was an outer layer of 2~mm of aluminium, with an inner layer of 0.5~mm of tantalum, with 1cm$\times$1cm$\times$1cm fused silica radiators. Each radiator had a simulated SiPM detector attached to one transparent side of the radiator. The two geometries are nearly identical, with the only difference being that the coincidence case has two Cherenkov radiators and simulated SiPM detectors, rather than just one.}
    \label{fig:geometry}
\end{figure}

The main settings that were used in the GRAS simulations are shown in \cref{tab:GRASsettings}, and the geometries used in simulations are shown in \cref{fig:geometry}. 

\begin{table}
\centering
\caption{The settings for GRAS\cite{santin2005gras} that were used in the simulations described in this work.}
\label{tab:GRASsettings}
\footnotesize
\begin{tabularx}{\textwidth}{ |X|X| } 
 \hline
 \textbf{GRAS Setting} & \textbf{Value} \\ 
 \hline
 \texttt{/gras/physics/addPhysics} & \texttt{em\_standard, G4OpticalPhysics, elastic, binary\_had, binary\_ion, decay, stopping, gamma\_nuc, raddecay} \\
\texttt{/process/optical/processActivation Cerenkov} & \texttt{True} \\
\texttt{/process/optical/processActivation Scintillation} & \texttt{False} \\
 \hline
\end{tabularx}
\end{table}

Either one or two 1cm$\times$1cm$\times$1cm cubic fused silica radiators were used, situated adjacent to each other in the two radiator case. One surface of each radiator was set to be transparent and correspond to where a SiPM detector \cite{gola2019nuv} would be in the real system. The refractive index of fused silica was set to be between 1.187 and 1.538 \cite{malitson1965interspecimen} as a function of wavelength, and light absorption lengths were set to be between $8.08 \times 10^{-5}$~cm and $82.2$~cm as a function of photon energy in accordance with Colvin et al.\cite{Colvin2011}. An outer shell of 2~mm of aluminium and an inner shell of 0.5~mm of tantalum were also placed surrounding the radiators. These shielding layers are intended to reduce count rates due to trapped electrons, and are also designed to act as a graded-Z shield in configuration to reduce count rates due to fluorescence.

Analysis was performed using a combination of the `Fluence' module for GRAS and Python post-processing. Every time a charged particle with a velocity exceeding the Cherenkov threshold passed through the fused silica, a number of Cherenkov photons would be generated along the track. One surface of the fused silica was set to be transparent - the surface corresponding to the SiPM device - whereas all other surfaces were set to be 99\% reflective for all photons. When a photon would pass through the SiPM surface, it would be tallied in the Fluence module and outputted to a file. Then in post-processing the photon detection efficiency of the SiPM device was applied to randomly reject photons in accordance with their wavelength, to simulate the actual observed count rates. The photon detection efficiency chosen was a 10~V overvoltage for a NUV-HD 40 µm cell shown in Figure 3 of Gola et al.\cite{gola2019nuv}.

In a real detector it should be possible to divide the photon count detection into multiple channels for different magnitudes of pulse heights/photon counts. However for the analysis presented in this article, it was assumed that there was only a single integral photon channel so that analysis was simplified. The default lower photon count threshold was taken to be 20 photons (the threshold was chosen somewhat arbitrarily as a threshold that might be used if the real-world detector experiences significant detector noise at lower photon count channels, but in real scenarios it is possible to even do single photon counting with SiPMs in Cherenkov detectors, e.g. Sottile et al. \cite{sottile2013uvsipm}), although it was possible to examine different photon count thresholds in post-simulation analysis too, and led to some interesting results that are described later in this article.

Output results had to be normalised to incoming isotropic spectra using calculated exposure times that could be calculated from\cite{fioretti2012low,lotti2021review,miller2022mitigating}: 

\begin{equation}
    T = \frac{N}{\Phi \times \pi \times A_G} = \frac{N}{\Phi \times 4\pi^2 R_G}
\end{equation}

where $T$ is the exposure time represented by a given simulation, $N$ is the number of incident particles simulated, $\Phi$ is the total integrated particles flux of the spectrum used in the simulation, in units of cm$^{-2}$ s$^{-1}$ sr$^{-1}$, and $A_G$ and $R_G$ are the area and radius of the sphere that particles were generated from in simulations. 

\begin{figure}
    \centering
    \includegraphics[width=\textwidth]{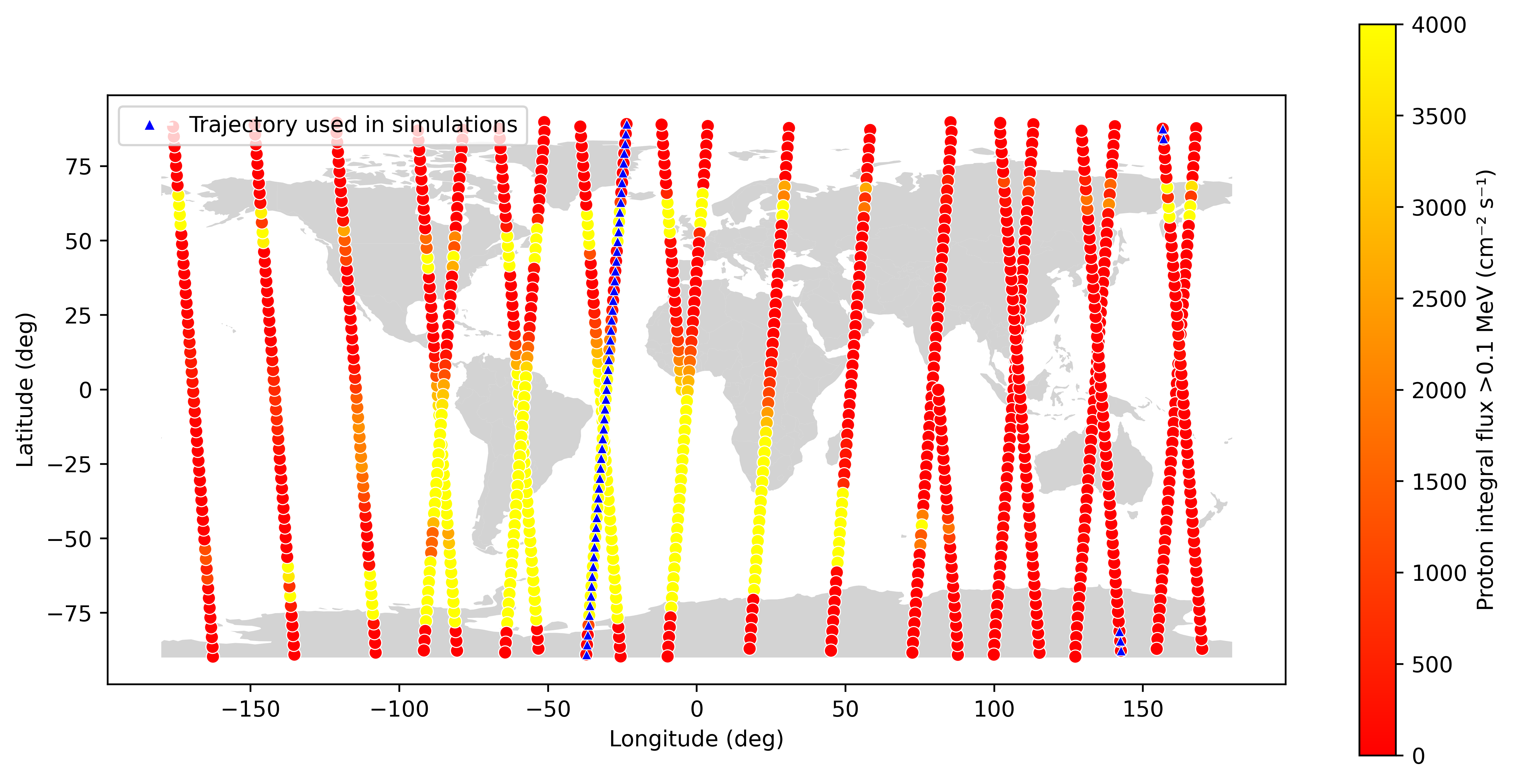}
    \caption{The 450~km altitude circular orbital trajectory that was simulated in this work is shown as the blue dotted line. This is overlayed on top of points that are coloured in accordance with simulated AP-8 $>$0.1~MeV proton fluxes, to show roughly where the horns regions and South Atlantic Anomaly (SAA) are. The Cherenkov detector system investigated here will actually likely only be affected by protons with kinetic energies much larger than 0.1~MeV, however the $>$0.1~MeV proton flux map shows the locations of the relevant radiation regions well, and shows that the investigated half-orbit passes through both the horns region and SAA region.}
    \label{fig:orbit}
\end{figure}

The representative half-orbit used as the coordinates for incoming spectra to be taken from is shown in \cref{fig:orbit}. The simulated trajectory was a 450~km altitude circular half-orbit that lasted 59.0 minutes. The half-orbit was generated using the SPENVIS orbit generator\cite{omaSPENVISSpace}, and passes through both the polar horns regions and the SAA region. As the half-orbit passes through the SAA, it could be considered a `maximum trapped particle half-orbit'.

\begin{figure}
    \begin{subfigure}{.5\textwidth}
        \centering
        \includegraphics[width=\linewidth]{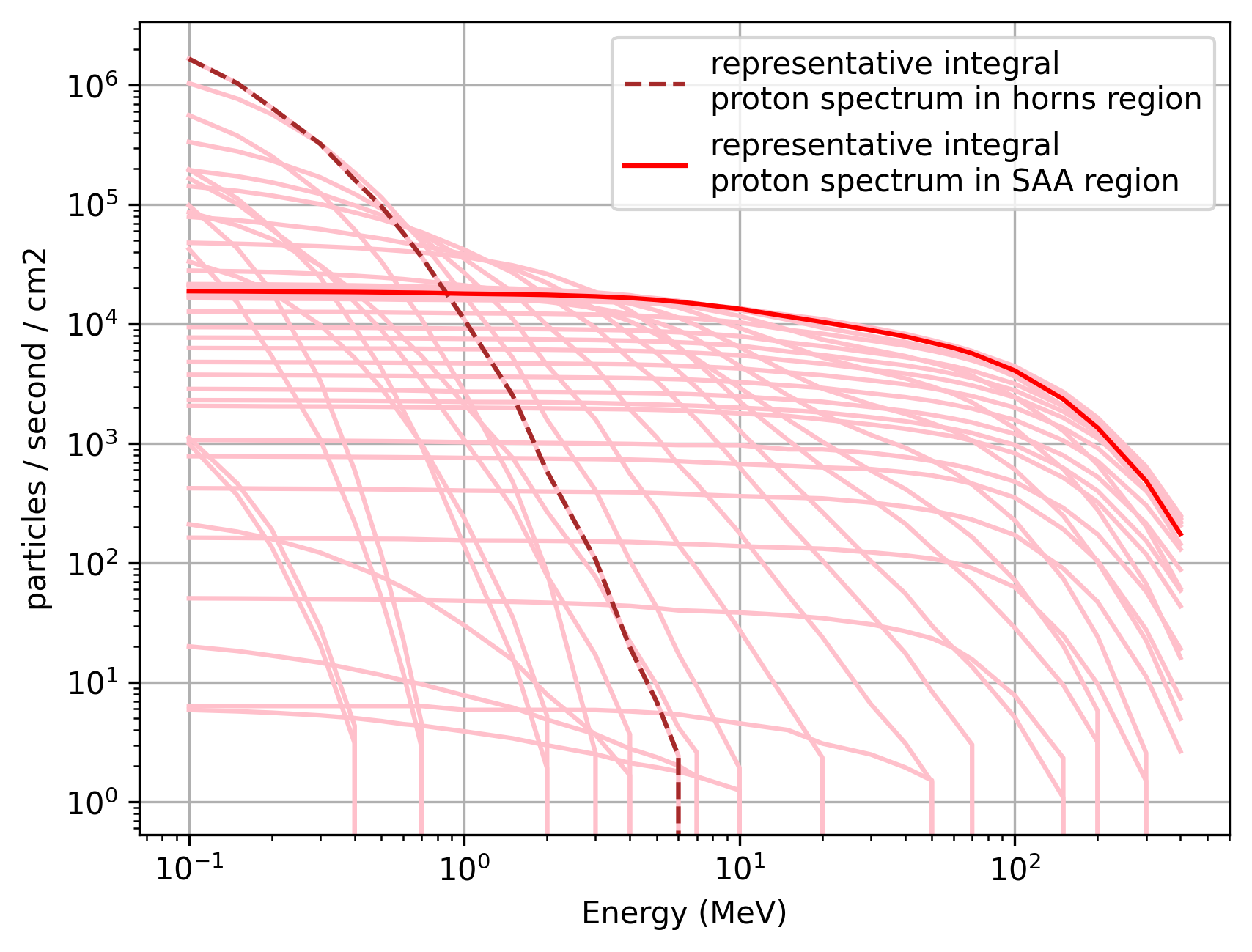}
        \caption{AP-8 trapped protons}
        \label{fig:proton_spectra}
    \end{subfigure}
    \begin{subfigure}{.5\textwidth}
        \centering
        \includegraphics[width=\linewidth]{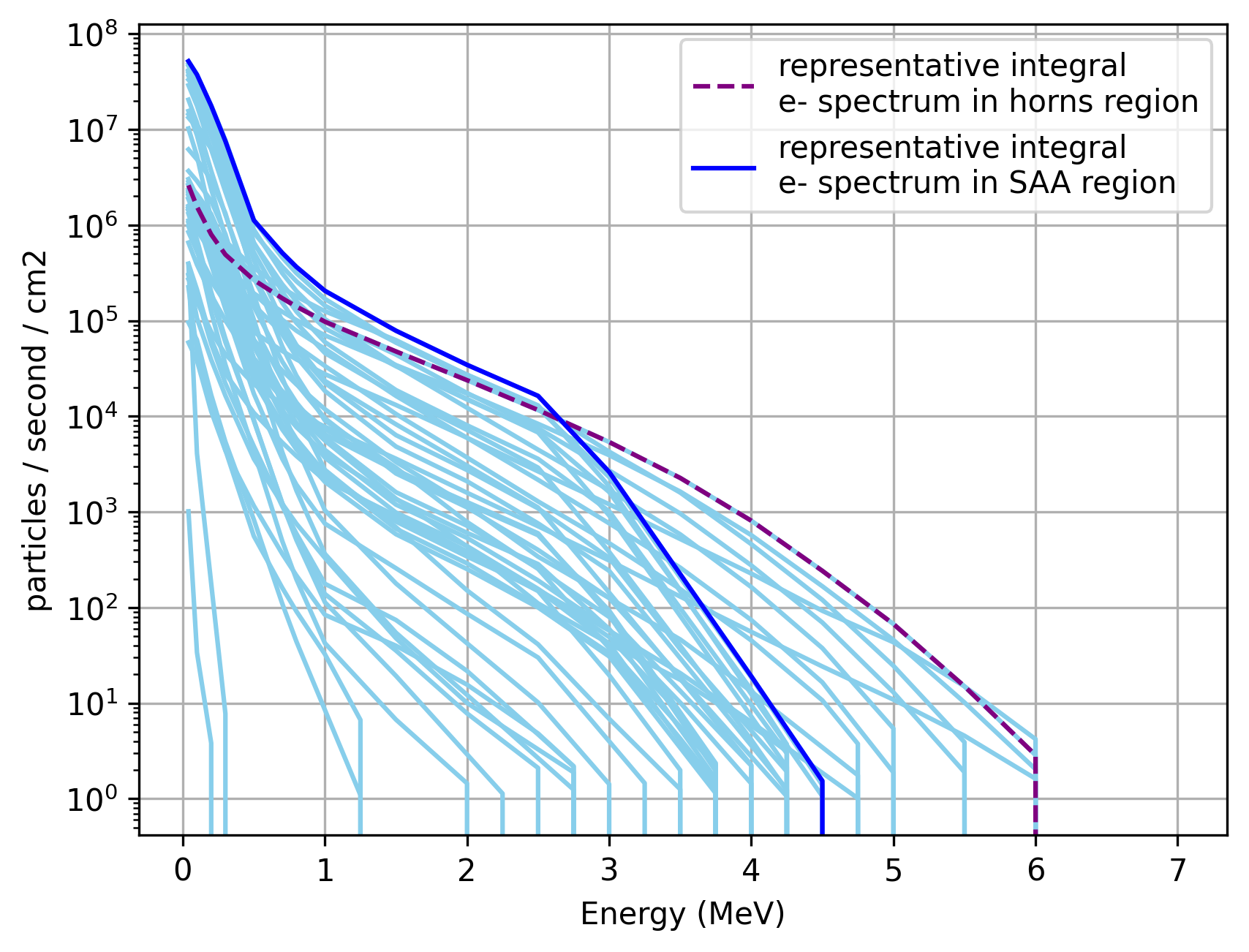}
        \caption{AE-8 trapped electrons}
        \label{fig:electron_spectra}
    \end{subfigure}
    \caption{All of the trapped particle spectra from AP-8 and AE-8 across the half-orbit that were simulated in this work. The maximum trapped particle spectra in the horns regions and SAA region are highlighted in these plots, and these maximum flux spectra were used to calculate the representative count rates discussed in this paper.}
    \label{fig:trapped_spectra}
\end{figure}

This half-orbit was fed into AE-8 and AP-8 \cite{vette1991tremp,vette1991ae8,sawyer1976ap8}, also in SPENVIS, to generate the trapped particle electron and proton spectra for each coordinate, which are shown in \cref{fig:trapped_spectra}. The coordinates that were used as representative of the polar, horns and SAA regions are given in \cref{tab:representative_coords}.

\begin{table}
\caption{Coordinates that were used to represent various regions of the low Earth orbit space environment.}
\label{tab:representative_coords}
\centering
\begin{tabular}{ |c|c|c| }
 \hline
 \textbf{Location} & \textbf{Latitude (°N)} & \textbf{Longitude (°E)} \\ 
 \hline
 Polar region & $-87.7$ & $142.6$ \\ 
 Horns region & $-66.0$ & $324.6$ \\ 
 SAA region & $-23.1$ & $327.9$ \\ 
 \hline
\end{tabular}
\end{table}

\begin{figure}
    \centering
    \begin{subfigure}{.45\textwidth}
        \centering
        \includegraphics[width=\linewidth]{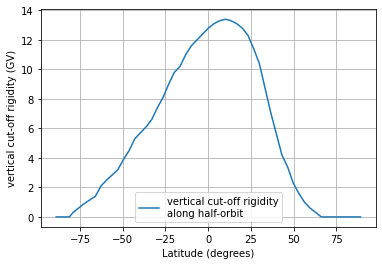}
        \caption{}
        \label{subfig:rigidity_used}
    \end{subfigure}
    \begin{subfigure}{.45\textwidth}
        \centering
        \includegraphics[width=\linewidth]{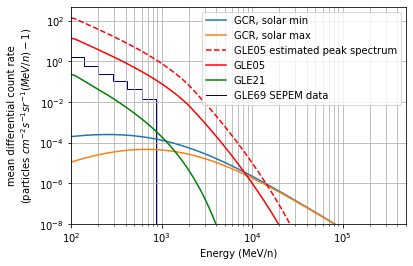}
        \caption{}
        \label{subfig:GLE_spectra}
    \end{subfigure}
    \caption{\Cref{subfig:rigidity_used} shows the calculated vertical cut-off rigidities that were used for the simulated trajectory in this study, as calculated according to MAGNETOCOSMICS\cite{MAGNETOCOSMICS} as implemented in the AsympDirsCalculator tool\cite{AsympDirsCalculator}. \Cref{subfig:GLE_spectra} shows the incoming interplanetary proton spectra that were simulated in this study.}
    \label{fig:GLE_spectra_details}
\end{figure}

A list of vertical cut-off rigidities were also generated for the half-orbit using MAGNETOCOSMICS\cite{MAGNETOCOSMICS} as implemented using the AsympDirsCalculator tool\cite{AsympDirsCalculator} for 1st January 2000, and are shown as a function of latitude in \cref{subfig:rigidity_used}. These vertical cut-off rigidities were used to truncate the incoming interplanetary particle spectra, which is a common approach used to model the effect of geomagnetic shielding on particle spectra at different latitudes and longitudes. 

The interplanetary spectra that were used in simulations are shown in \cref{subfig:GLE_spectra}. The GLE spectra used correspond to averaged spectra across GLE21 and GLE05 according to Koldobskiy et al.\cite{koldobskiy2021new}. GLE21 and GLE05 were chosen as representative SEP events to represent a small event ($\sim$10\% maximum neutron monitor increase) and a large event ($\sim$4,600\% maximum neutron monitor increase), respectively\cite{GLEdatabase, usoskin2015database}.

The galactic cosmic proton spectrum used corresponded to solar maximum at the 1st of January 2000 according to the CosRayModifiedISO Python package \cite{davis_cosraymodifiediso_nodate}, which uses a variant of the ISO model, modified by DLR\cite{matthia2013ready}. The galactic cosmic proton spectrum therefore corresponds to a minimum flux of galactic cosmic protons.

\section{Count Rates due to Trapped Particles, and Discussion}
\label{sec:Simulation_Results_Trapped}

\begin{figure}
    \begin{subfigure}{.5\textwidth}
        \centering
        \includegraphics[width=\linewidth]{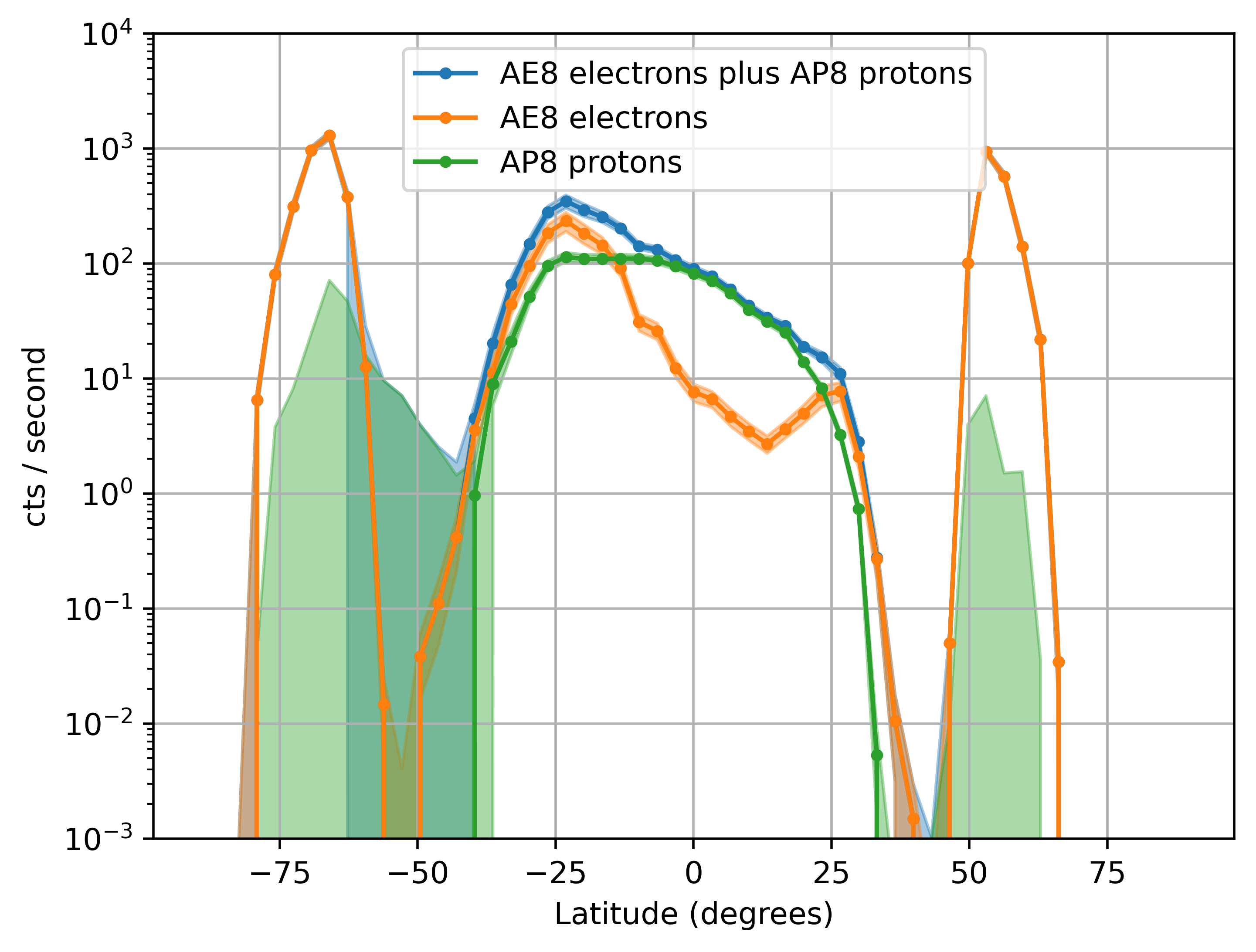}
        \caption{No coincidence}
        \label{subfig:single_trapped_cr}
    \end{subfigure}
    \begin{subfigure}{.5\textwidth}
        \centering
        \includegraphics[width=\linewidth]{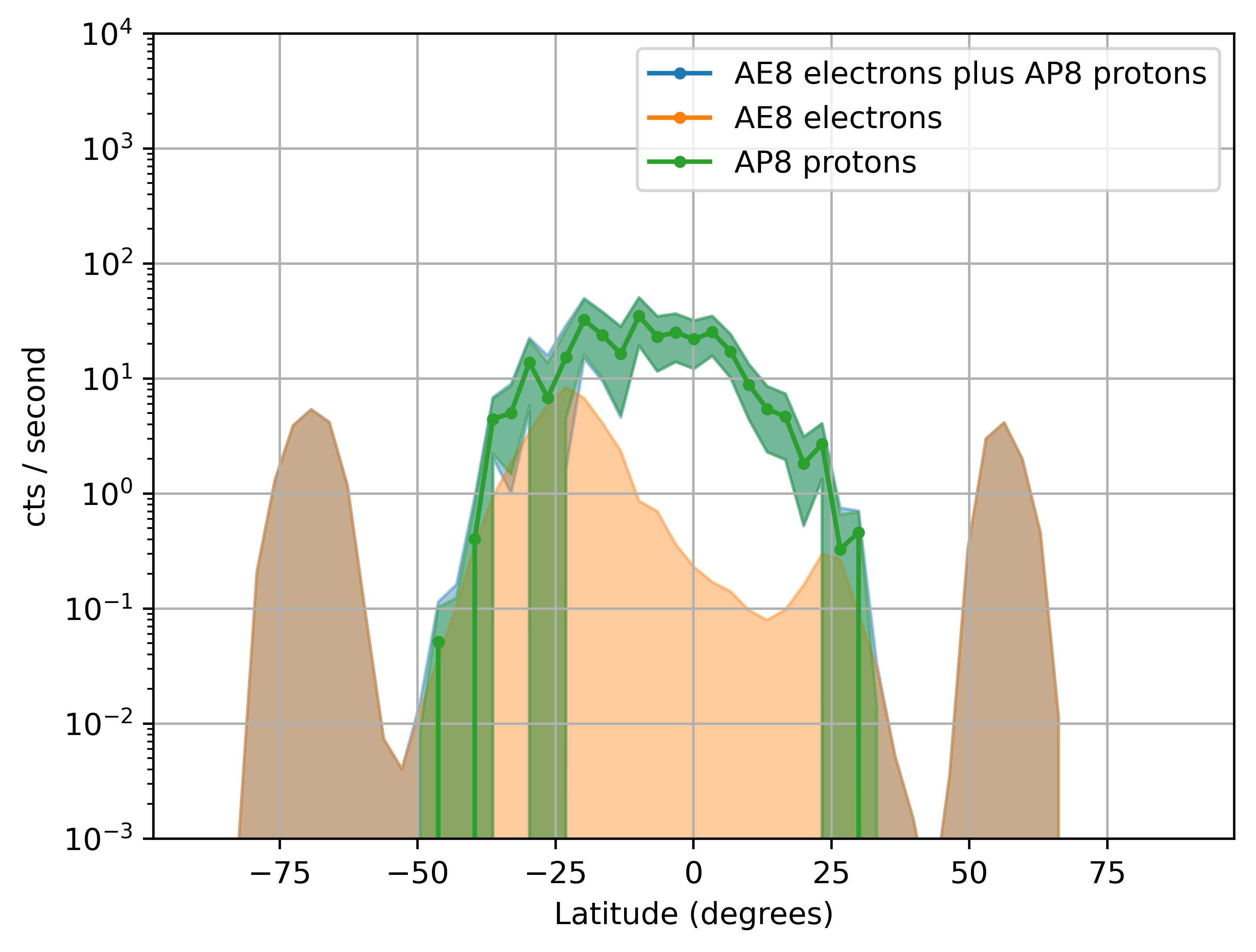}
        \caption{With coincidence}
        \label{subfig:coincidence_trapped_cr}
    \end{subfigure}
    \caption{Observed count rates due to trapped particles as a function of latitude along the half-orbit for both the single radiator case, and the dual radiator coincidence case. Here the shaded regions exclusively represent error bars (i.e. in \cref{subfig:coincidence_trapped_cr}, no electrons were detected but the simulation statistics mean that error bars are still present up to a maximum of 10 cts/s. Where no counts were detected, error bars were approximated as corresponding to a single particle detection). In the single radiator case, both the horns region and south Atlantic anomaly region show high count rates due to trapped particles, which would contribute a significant background for making observations of other particle types. The coincidence case successfully removes the trapped particle components from the horns region, and reduces trapped particle count rates from SAA significantly, but not entirely. Note that in the coincidence case, count rates from the horns region, confidence limits and error bars were excluded. This is because high statistics (albeit time-consuming) individual point simulations showed that trapped particle count rates in this region were less than 1 cts / second (and likely significantly less than this).}
    \label{fig:trapped_cr}
\end{figure}

The total integral count rates are shown in \cref{fig:trapped_cr}, for the single radiator and coincidence mode cases. As you would expect, the count rates peak in both the horns regions and in the South Atlantic anomaly region. 

In the single radiator case, both trapped protons and trapped electrons are important for count rates in the South Atlantic anomaly region (here between -50 degrees and 30 degrees in latitude). Much of the SAA is dominated by a somewhat continuous flux of trapped protons, while there are two trapped electron peaks at about -25 degrees and 25 degrees in latitude. In contrast, the horns region is dominated exclusively by electrons. Separate individual coordinate simulations utilising a high number of simulated protons indicated that there are less than 0.71 cts / s (possibly significantly less) due to trapped protons in the horns region. These count rates are very visible over short time periods, reaching over 10 cts per second across the majority of the trapped particle orbital length, meaning that even this relatively small Cherenkov detector should be readily able to perform scientific measurements of the van Allen belt populations.

In the coincidence mode case, trapped electrons count rates are almost entirely removed, and possibly removed completely. Separate individual coordinate simulations indicated that there are $<$0.42 cts / s due to trapped electrons in the horns region when coincidence is applied. In the SAA region trapped proton-induced count rates are reduced significantly, however not entirely, and a significant quantity of trapped protons remains visible, reaching a maximum of 14.4$\pm$3.7~cts~/~s at the position of the single radiator SAA maximum. The coincidence mode here could therefore be used within the trapped particle context to perform observations of specifically the trapped proton populations of the SAA, while removing trapped electrons.

\begin{figure}
    \begin{center}
    \begin{subfigure}{.5\linewidth}
        \centering
        \includegraphics[width=\linewidth]{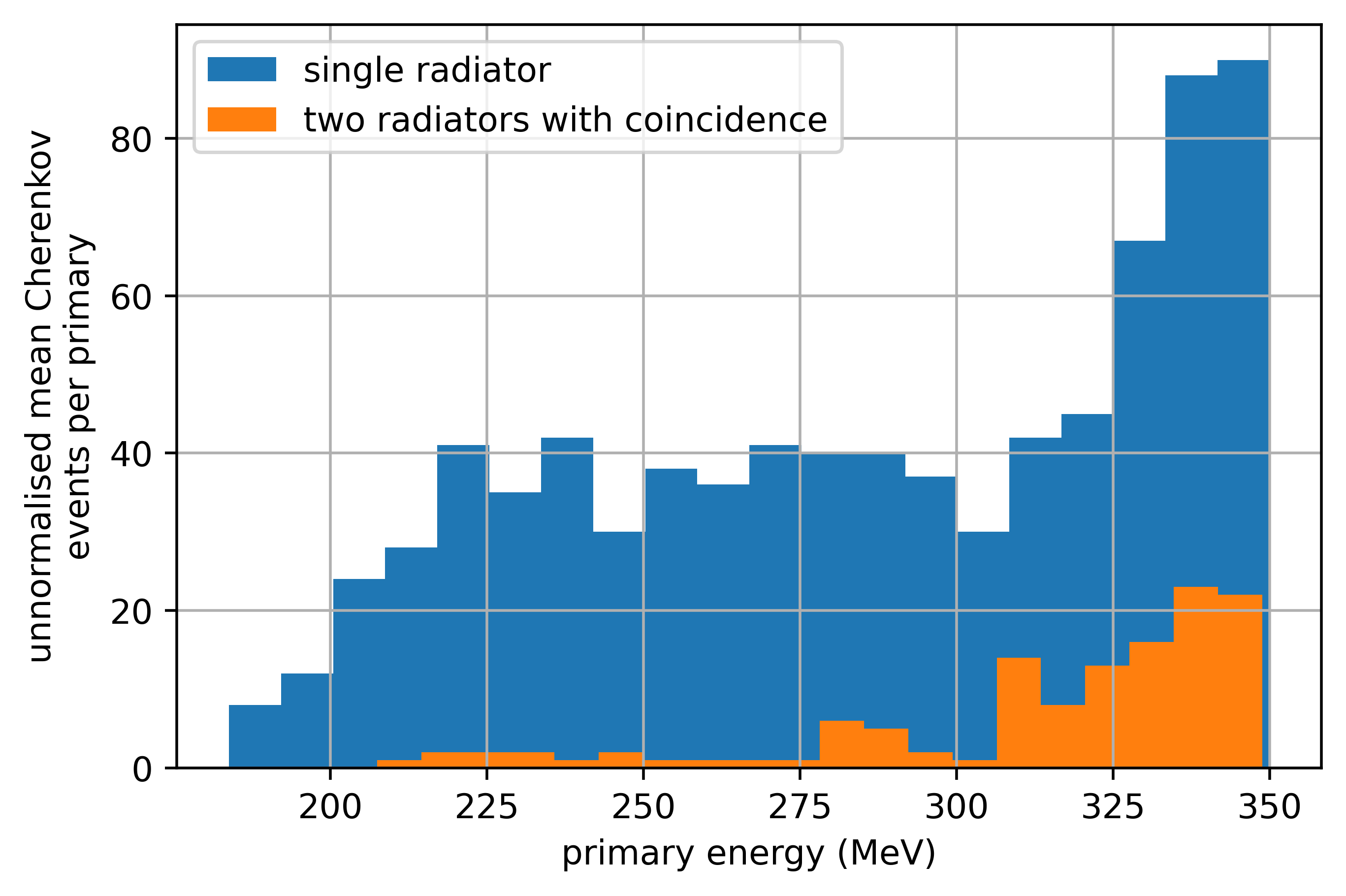}
        \caption{}
        \label{subfig:delta_electron_origin}
    \end{subfigure}
    \end{center}
    \begin{subfigure}{.5\textwidth}
        \centering
        \includegraphics[width=\linewidth]{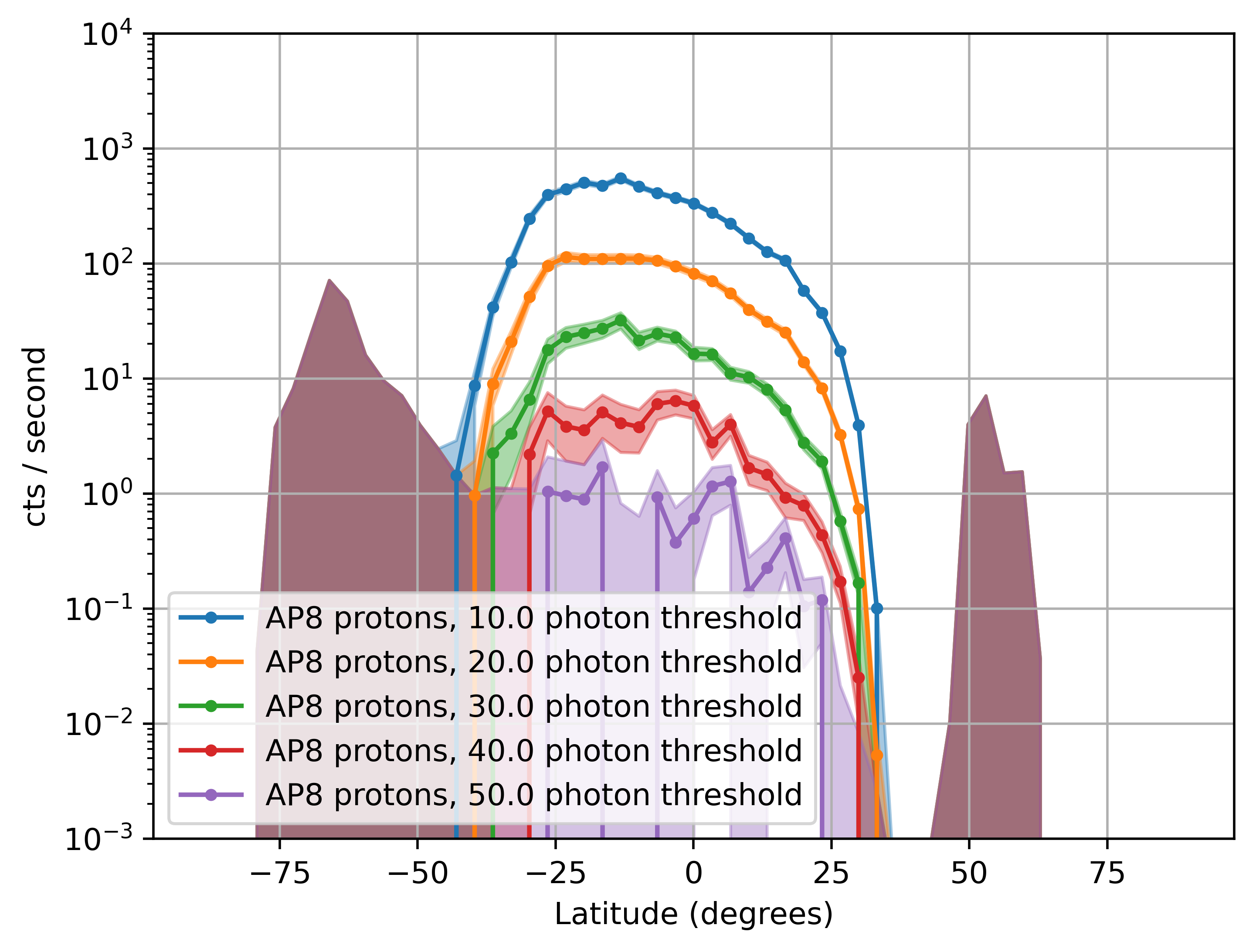}
        \caption{}
        \label{subfig:protons_with_thresholds}
    \end{subfigure}
    \begin{subfigure}{.5\textwidth}
        \centering
        \includegraphics[width=\linewidth]{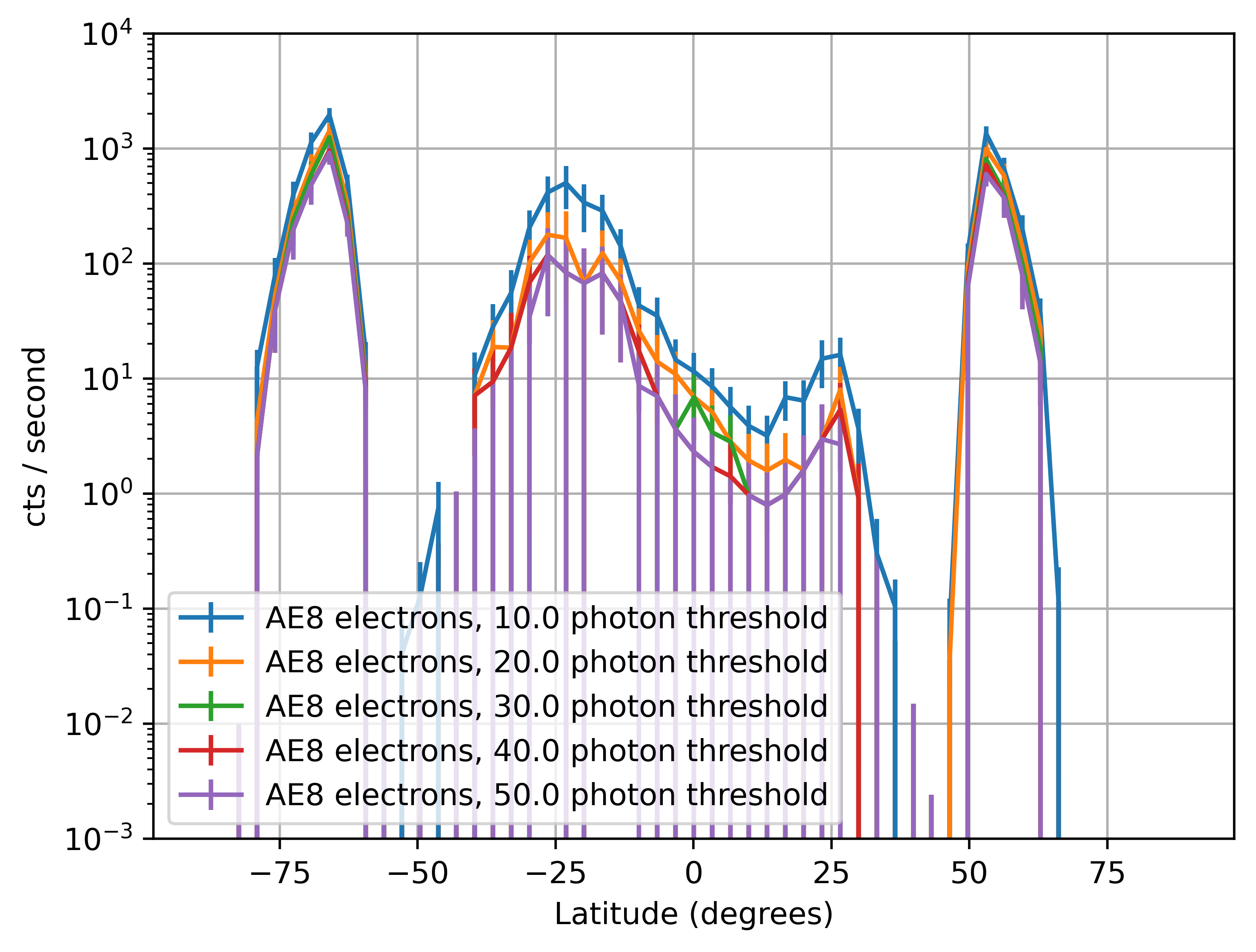}
        \caption{}
        \label{subfig:electrons_with_thresholds}
    \end{subfigure}
    \caption{\Cref{subfig:delta_electron_origin} shows the distribution of primary particle kinetic energies that induced a valid detectable event in the detector for trapped protons in the SAA region in the single radiator and coincidence case, and shows that there is a component both above the Cherenkov threshold (which is a minimum of 296.7~MeV for fused silica, for the maximum refractive index of 1.538 simulated here), and below the Cherenkov threshold. \Cref{subfig:protons_with_thresholds,subfig:electrons_with_thresholds} shows how the trapped particle count rates in the single radiator case varies when the photon threshold for a valid event is varied from the default 20 detected photons, showing a significant variation.}
    \label{fig:trapped_cr_primary_energies}
\end{figure}

It was surprising to us that there was such a strong signal associated with trapped protons, and it was instructive to investigate this further. The count rates associated with trapped protons as a function of trapped proton kinetic energy are shown in \cref{subfig:delta_electron_origin}, and as a function of photon threshold in the single radiator case in \cref{subfig:protons_with_thresholds}.

\Cref{subfig:delta_electron_origin} indicates that the count rate due to trapped protons originates from both trapped protons below and above the Cherenkov kinetic energy threshold for protons (between 296.7~MeV and 800.9~MeV in fused silica for the refractive indices used in these simulations). The likely mechanism for count rates due to protons below the Cherenkov kinetic energy threshold is the generation of delta electrons in the fused silica as a proton of several hundred MeV passes through it. The range of primary energies below the Cherenkov threshold that are represented in \cref{subfig:delta_electron_origin} corresponds well with the delta electron response shown in figure 2 of Mazur et al.\cite{mazur2023relativistic}. 

The median energy of particle that directly induced a photon (i.e. the parent particle of the photon) with energies less than the proton Cherenkov threshold range was 0.311~MeV. The 25th and 75th percentiles were 0.261~MeV and 0.376~MeV, respectively. These values align well with the electron Cherenkov threshold in fused silica, which ranges from 0.162~MeV to 0.436~MeV for the refractive indices used in these simulations. It is also interesting that the coincidence mode was not entirely able to remove this delta electron component, despite reducing it significantly, which might be expected for a fully stochastic delta electron generation process in each radiator.

Another interesting feature shown in \cref{subfig:delta_electron_origin} is that there appear to be a significant quantity of counts generated above the Cherenkov threshold too, showing that AP-8 suggests there are a lot of high energy trapped protons in the SAA region. The maximum energy for AP-8 is in fact the 350 MeV limit on the figure, and the fact that the count rates are increasing up until that point and presumably beyond it shows that count rates due to trapped protons may actually be underestimated in this analysis, assuming that AP-8 is correct. The ability to investigate these trapped proton populations using the Cherenkov designs discussed in this paper would be useful scientifically and indeed be possible as \cref{subfig:protons_with_thresholds} shows that the count rates due to them varies significantly with photon threshold/photon channel. 

\Cref{subfig:electrons_with_thresholds} also shows that trapped electron count rates vary somewhat with photon threshold too, although less significantly and it is likely analysis methods such as spectral deconvolution could be developed to convert photon distributions due to these trapped particles into a spectra for both particle species.

\section{The Effect of Trapped Particle Background on Detection of Galactic Cosmic Protons and Solar Energetic Protons}
\label{sec:Simulation_Results_Cosmic_SEPs}

\begin{figure}
    \begin{subfigure}{.5\textwidth}
        \centering
        \includegraphics[width=\linewidth]{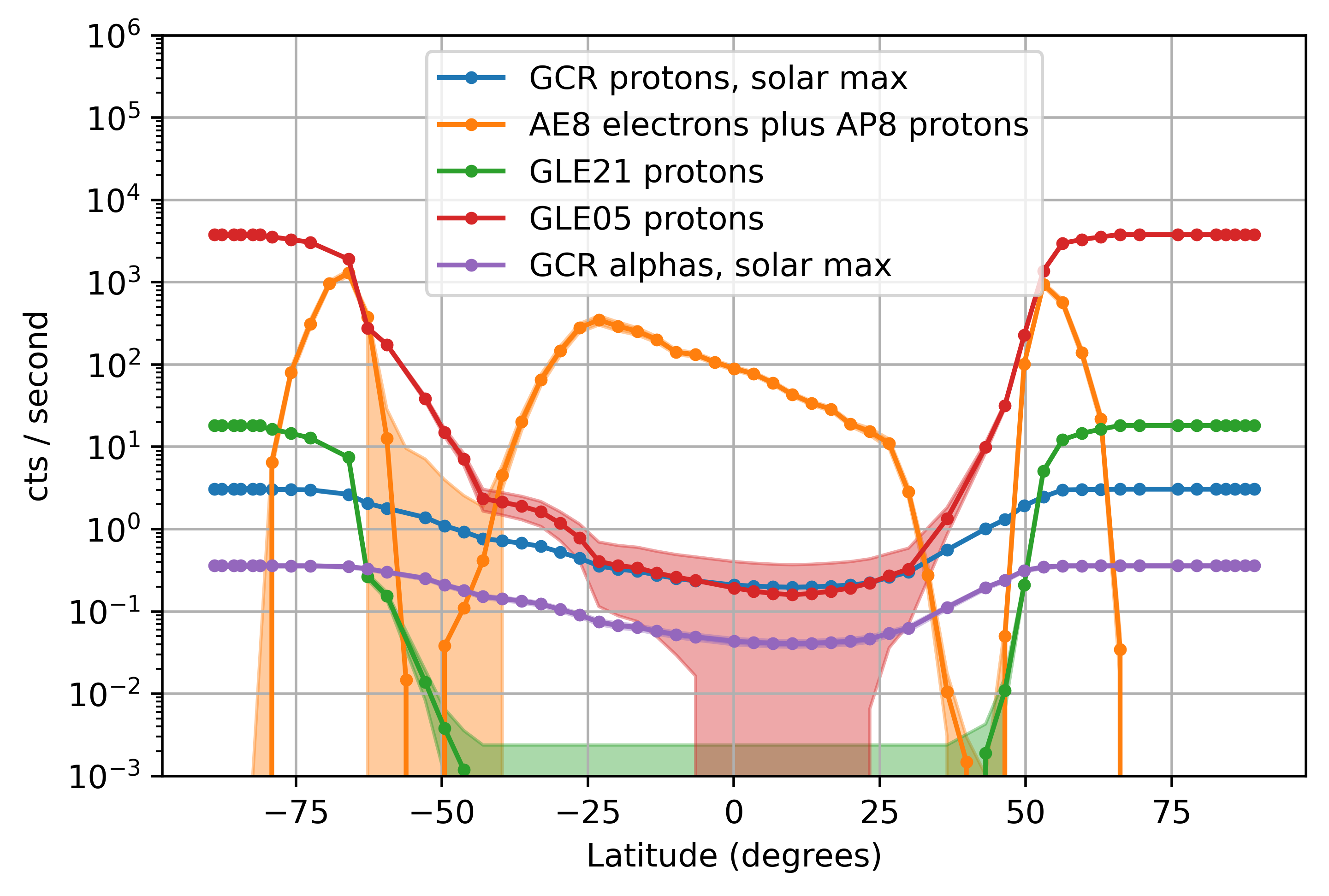}
        \caption{No coincidence}
        \label{subfig:GLE_GCR_cr}
    \end{subfigure}
    \begin{subfigure}{.5\textwidth}
        \centering
        \includegraphics[width=\linewidth]{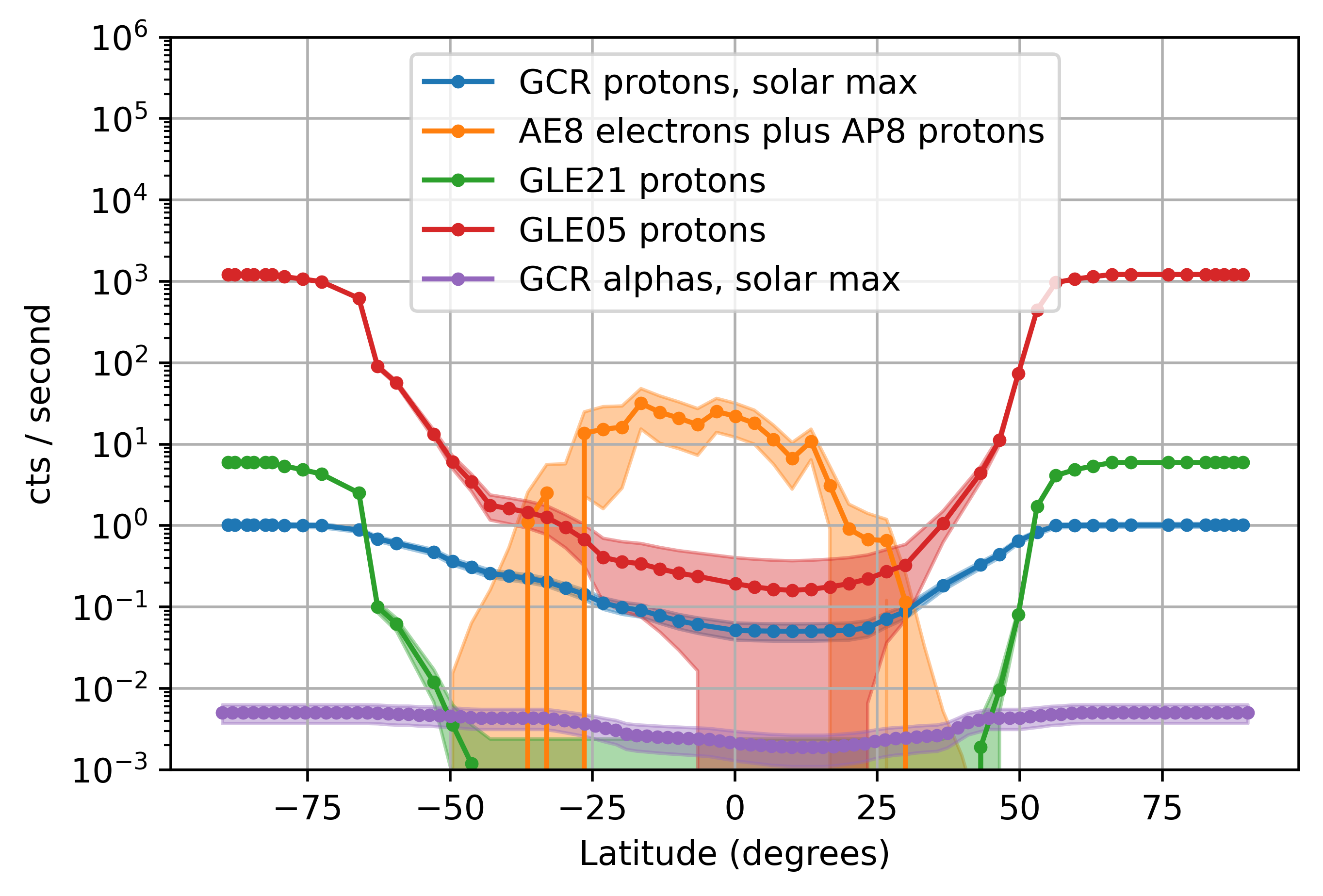}
        \caption{With coincidence}
        \label{subfig:GLE_with_thresholds}
    \end{subfigure}
    \caption{Count rates due to all the simulated particle components across the simulated half-orbit, for both the single radiator case, and the multi-radiator/coincidence case. Count rates due to cosmic rays and GLEs all reduce towards equatorial regions due to increased cut-off rigidities albeit more significantly for softer incoming proton spectra. The coincidence mode reduces count rates due to these interplanetary proton spectra by two thirds, while reducing trapped particle count rates much more significantly, potentially removing them completely in the horns regions and by a factor of $\approx22$ at the peak of the SAA region (a latitude of $-23$\textdegree~for this half-orbit).}
    \label{fig:all_components_cr}
\end{figure}

Count rates as a function of latitude for each of the different energetic proton sources are plotted in \cref{fig:all_components_cr}. The total count rate due to trapped particles is also plotted here. \Cref{subfig:GLE_GCR_cr} indicates that the single radiator count rates due to SEP protons and GCR protons in the SAA region and horns regions would most likely be obscured. 

When coincidence mode is applied, the horns region becomes completely visible for each of the GLEs and GCR case. Note that the horns region would be particularly important for observations of GLE protons, because the horns region roughly follows geomagnetic latitudes, and would therefore permanently exclude observations of particles at the particular horns region vertical cut-off rigidities. The SAA region in the coincidence case still remains intense due to trapped protons, as was discussed previously, which could exclude observations of particularly high energy GLEs at equatorial latitudes. This would not be a major problem at 450~km in altitude as simulated here, as most orbits would miss the SAA, however at higher altitudes this could be more of a problem as the SAA grows larger in size.

As with the trapped proton and trapped electron case, it is likely possible to perform spectral reconstruction using photon channels as well as using vertical cut-off rigidities, to get true in-situ spectral measurements. However the issue of delta electrons induced by incoming particles could create challenges with applying analytical deconvolution methods to a range of photon channels, which would normally rely on the assumption that all photons are induced by a single particle species.

While these simulations describe the count rates due to the expected dominant components, several simulations were also performed examining the gamma components. Simulations of the cosmic X-ray background (using a spectrum given by Moretti et al.\cite{moretti2012spectrum}) gave a count rate of $0.08\pm0.01$ per second. Simulations were also performed using a polar spectrum of albedo gamma rays (using the approach given in Murphy et al.\cite{murphy2021compact}), which gave a count rate totalling $16.1\%\pm0.9\%$ of the count rates associated with a polar region GCR proton spectrum. While individually these components are not significant, they could collectively sum to contribute a reasonable amount to count rates, at least during quiet solar conditions, although the gamma components will almost certainly be removed in the coincidence case (and anticoincidence could also potentially be used to measure these components).

One component that was not simulated in this paper were cosmic electrons, as before the simulations were being built and run it was assumed that they would be too low in intensity and too well shielded by Earth's magnetosphere and spacecraft shielding to be significant along the half-orbit. However, subsequent back-of-the-envelope calculations reveal that cosmic electrons could become significant in geomagnetic polar regions: a 10~MeV electron is able to penetrate 22~mm of shielding \cite{berger2005estar}, corresponding to roughly the thickness of aluminium used in this paper (well above the electron Cherenkov threshold of 0.162~MeV for the maximum refractive index of 1.538 for the fused silica simulated here). Integrating the VLIS electron spectrum given by Potgieter and Vos \cite{PotgieterDifference2017} from 10~MeV to infinity gives a total integral flux of 0.638~cm$^{-2}$sr$^{-1}$s$^{-1}$. Integrating the Potgieter and Vos cosmic proton spectrum from \>300 MeV (corresponding approximately to the fused silica Cherenkov threshold) gives a similar value of 0.859~cm$^{-2}$sr$^{-1}$s$^{-1}$, implying that While cosmic electrons will be blocked by even minimal geomagnetic shielding (for example the magnetic rigidity of a 50~MeV electron is only 0.00549~GV), cosmic electrons could be visible even amongst cosmic proton count rates in geomagnetic polar regions, and it would be interesting for follow-on research to investigate this possibility. 

\begin{figure}
    \begin{subfigure}{.5\textwidth}
        \centering
        \includegraphics[width=\linewidth]{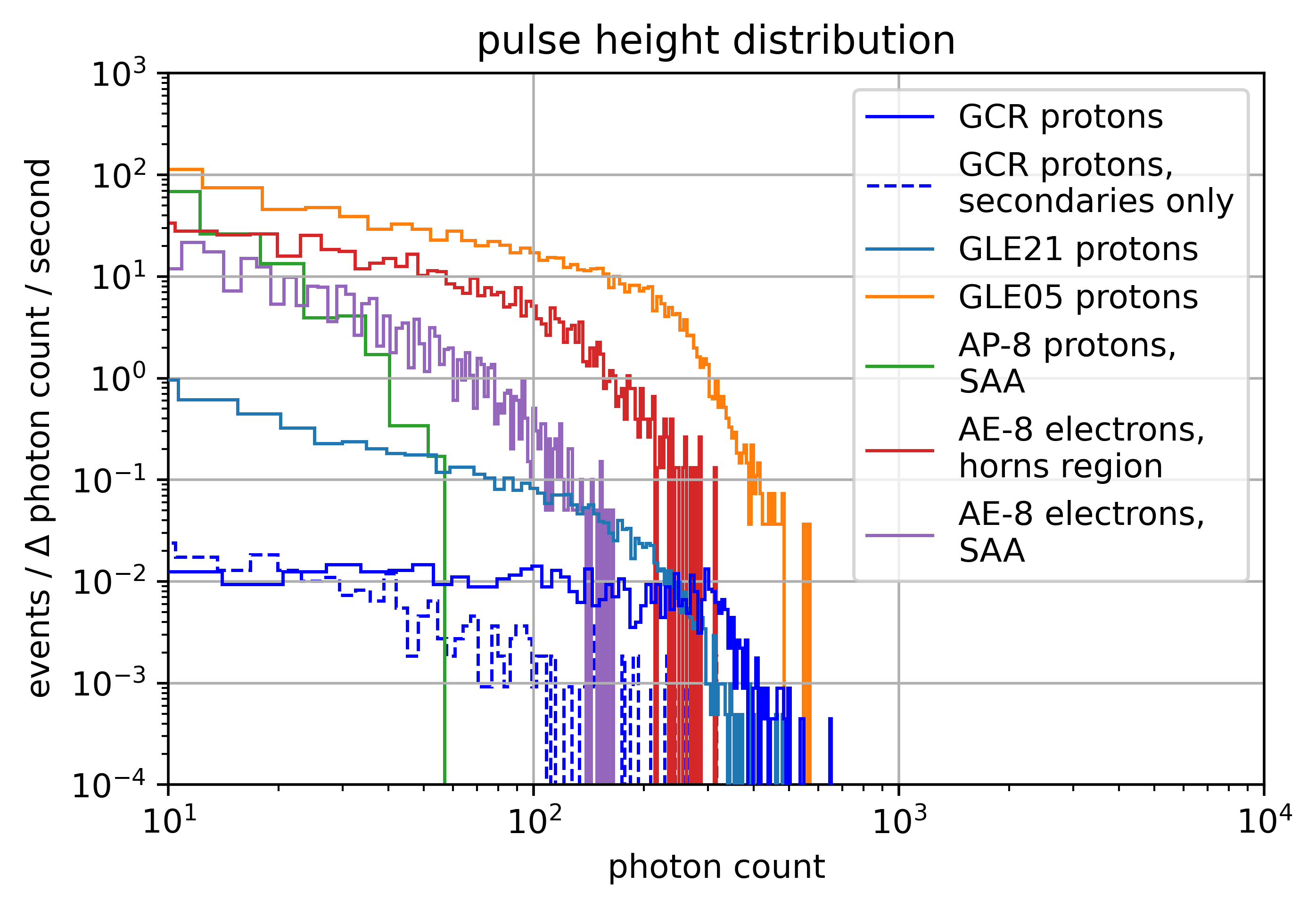}
        \caption{No coincidence}
        \label{fig:GLE_GCR_pulse_height_single}
    \end{subfigure}
    \begin{subfigure}{.5\textwidth}
        \centering
        \includegraphics[width=\linewidth]{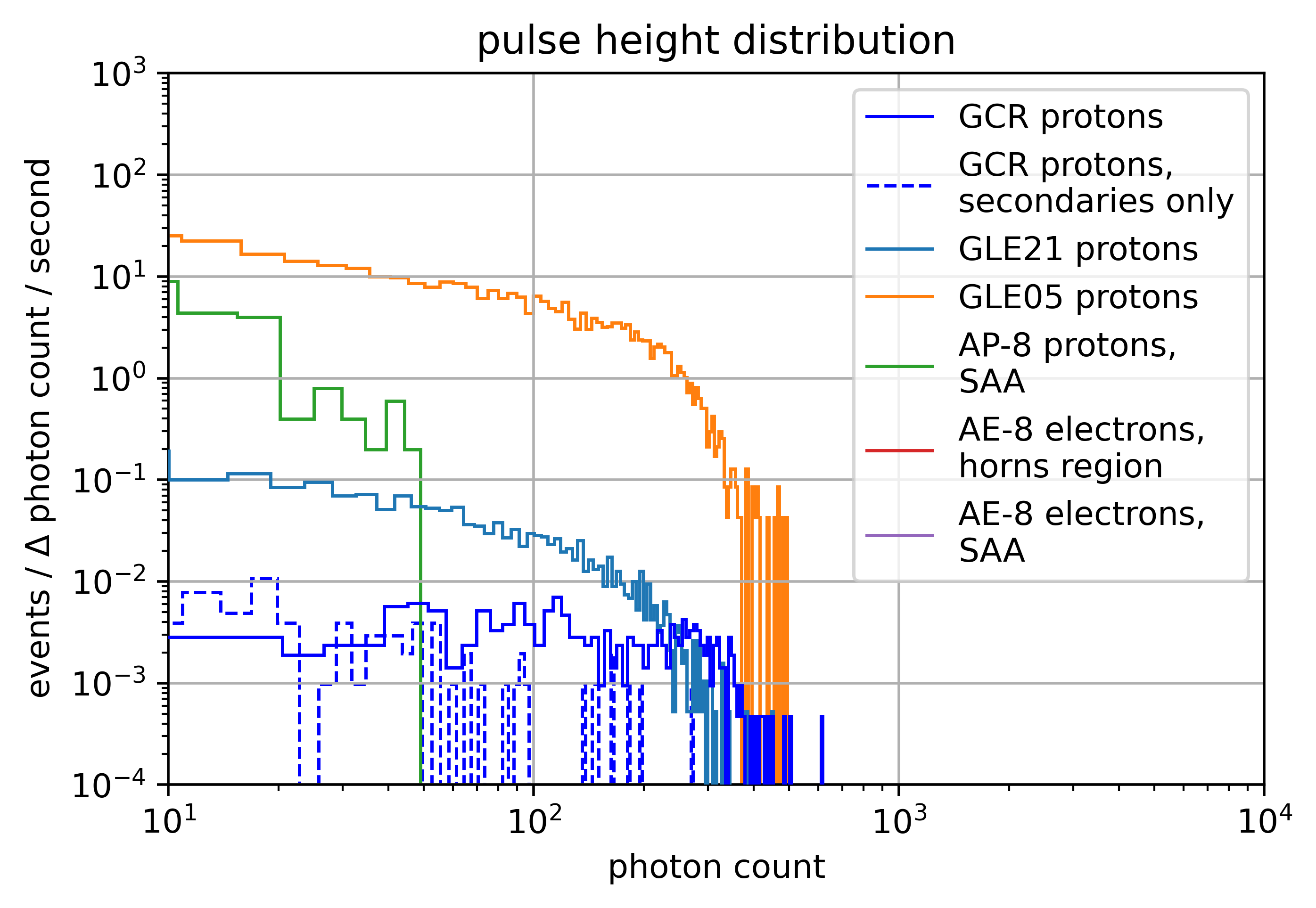}
        \caption{With coincidence}
        \label{fig:GLE_GCR_pulse_height_coincidence}
    \end{subfigure}
    \caption{The pulse height distributions for the trapped particle and interplanetary proton spectra.}
    \label{fig:GLE_GCR_pulse_height}
\end{figure}

As well as examining raw count rates that the detectors can observe, it was also very valuable to examine the actual pulse height distributions the detectors would observe. \Cref{fig:GLE_GCR_pulse_height} shows the photon count distributions when the full interplanetary proton spectra are incident upon the simulated system (no geomagnetic shielding) against photon count distributions when photons directly generated by the incoming primary are excluded. Note that in this case due to computational limitations for high statistics, only the dual radiator geometry was run, but the single radiator case was modelled in post-simulation analysis by just selecting outputs from a single radiator.

The pulse height distributions in \Cref{fig:GLE_GCR_pulse_height} each have different shapes and intensities, indicating that pulse height distributions could likely be used as a way to discriminate between different particles. The pulse height distribution due to galactic cosmic proton-induced secondary electrons is also shown, and is significant at low photon counts, while not appearing to be reduced significantly by the coincidence mode. The presence of secondary particle components that are naturally folded into the total pulse height distribution for the proton spectra would however complicate potential spectral reconstruction attempts using pulse height distributions alone.

The percentage of photons that are generated by secondary particles, when a lower exclusion threshold of 20 photon counts is applied, is given in \cref{tab:secondary_photon_percentages}, which ranges between about 5\% and 8\%. These percentages may certainly be large enough to introduce errors in any deconvolutions where only the presence of the primary proton is assumed, and therefore more complex methods may be required. Secondary particles were identified in these cases and in \cref{fig:GLE_GCR_pulse_height} by filtering for photons generated by parent particles with kinetic energies less than 200 MeV, using the RA module rather than the FL module. 

\begin{table}
\caption{The percentage of valid photons that were generated by secondary particles. The coincidence set-up only appears to remove a small fraction of these photons.}
\label{tab:secondary_photon_percentages}
\centering
\begin{tabular}{ |c|c|c| } 
 \hline
 \textbf{Incoming Spectrum} & \textbf{Single Radiator (\%)} & \textbf{Coincidence (\%)} \\ 
 \hline
 GCR protons & $6.50\pm0.06$ & $7.62\pm0.11$ \\ 
 GLE21 protons & $5.988\pm0.032$ & $5.75\pm0.05$ \\ 
 GLE05 protons & 5.461$\pm$0.018 & 5.269$\pm$0.030 \\ 
 \hline
\end{tabular}
\end{table}

Despite the fact that secondary particles will likely cause issues with deconvolution, it can be seen in \cref{fig:GLE_GCR_pulse_height} that the different radiation sources do generate quite different pulse height spectra (PHS) in the detectors. The difference between the GCR proton PHS, and the PHS for the two GLE spectra have very different shapes (for instance for a photon count of 79.8 the PHS of GLE05 is 209.6 times greater than that of GLE21, whereas at a photon count of 263.6, the PHS of GLE05 is 458.7 times larger), which indicates that spectral deconvolution of photon channels would likely be possible assuming background due to trapped particles and other sources is not significant.



\begin{figure}
    \begin{subfigure}{.5\textwidth}
        \centering
        \includegraphics[width=\linewidth]{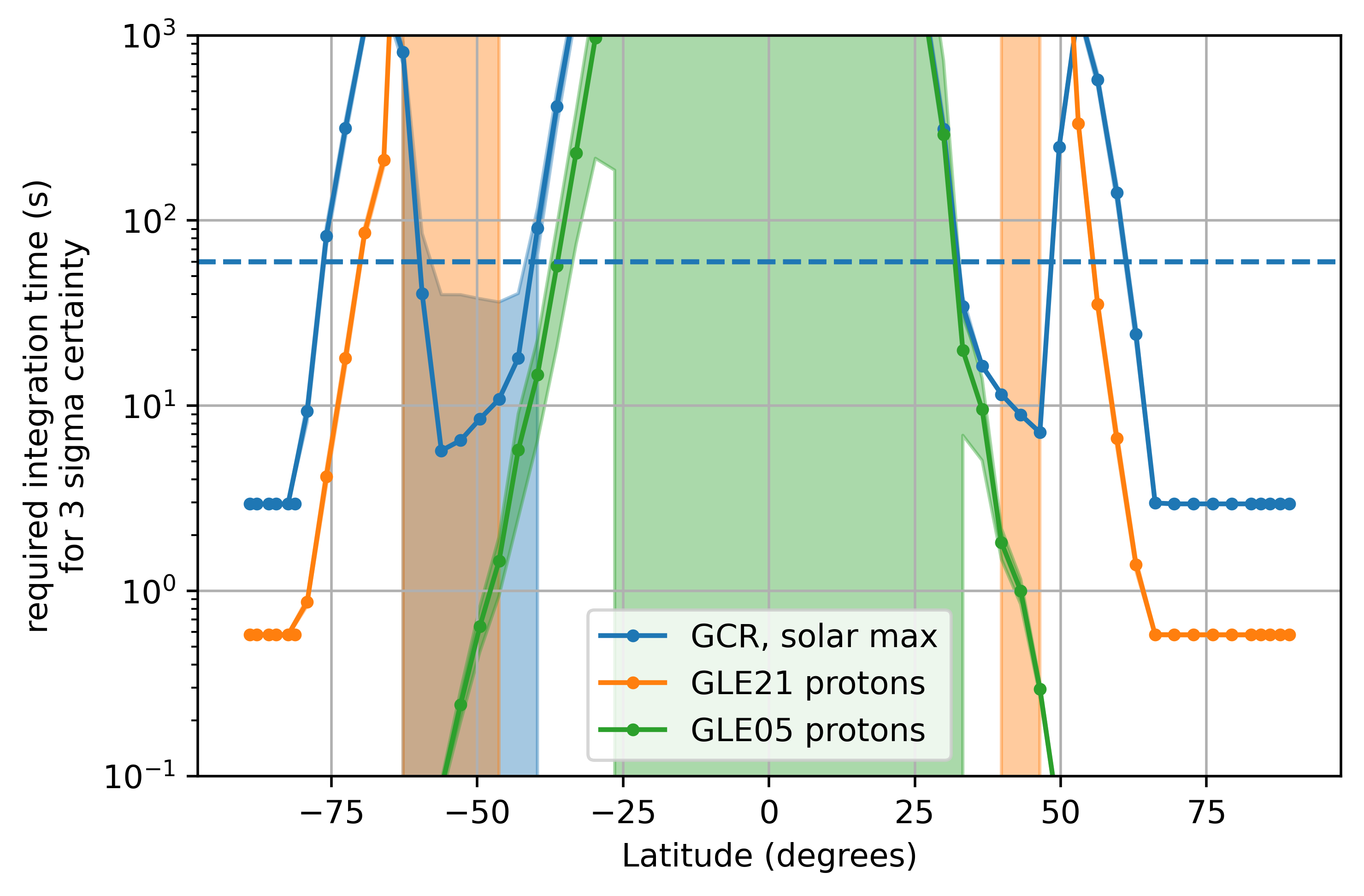}
        \caption{No coincidence}
        \label{fig:GLE_req_int_time}
    \end{subfigure}
    \begin{subfigure}{.5\textwidth}
        \centering
        \includegraphics[width=\linewidth]{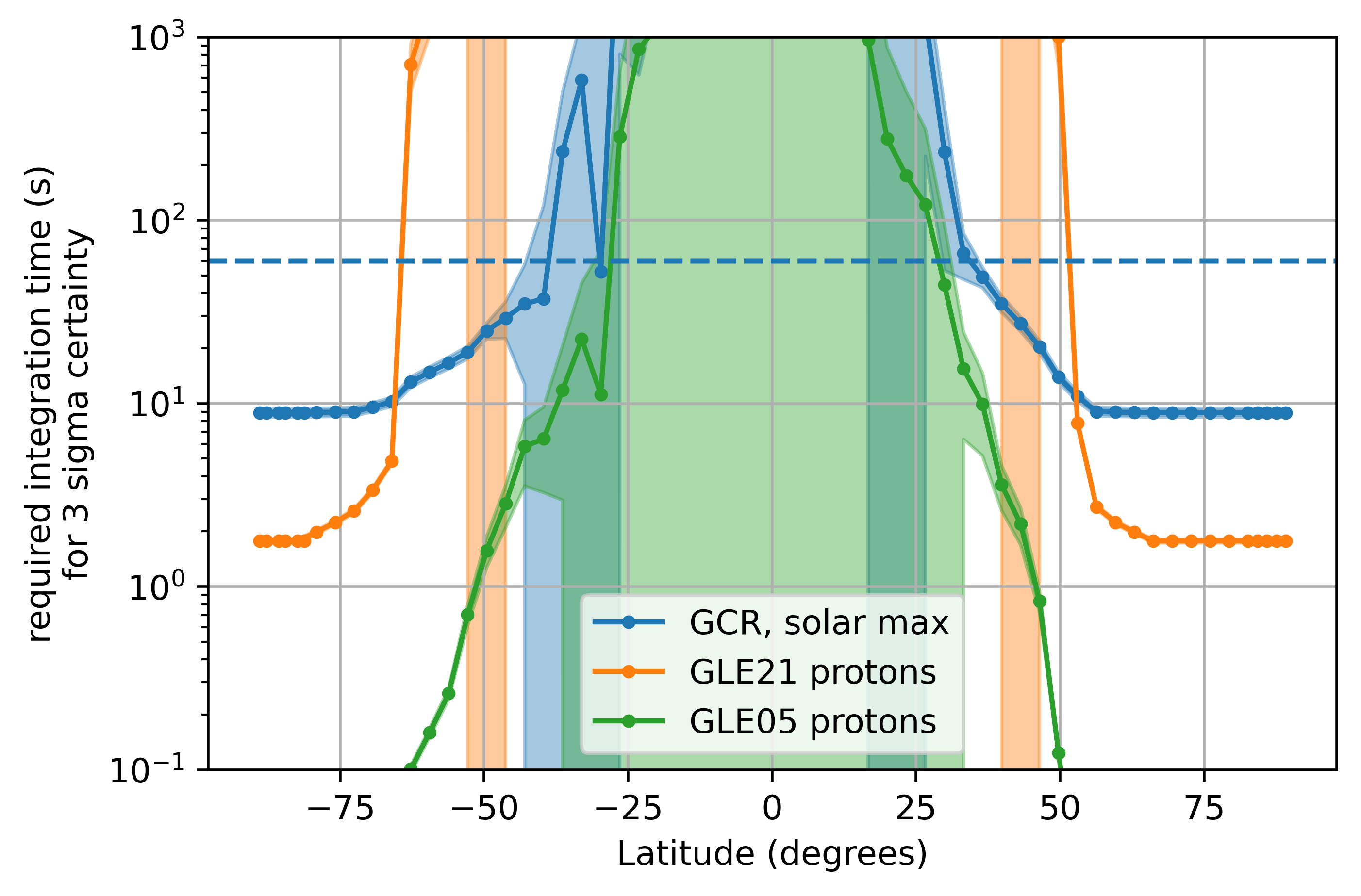}
        \caption{With coincidence}
        \label{fig:GLE_req_int_time_coincidence}
    \end{subfigure}
    \caption{The required integration times that would be needed to acquire count rates to a statistical significance of 3 sigma, when both Poisson noise and background count rates are taken into account (and where the mean background count rate is assumed to be accurately known).  A dashed line showing 1 minute in integration time is also plotted to show a rough limit for how low integration times might have be for a 59 minute half-orbit.}
    \label{fig:int_times}
\end{figure}

To further quantify the ability of the simulated geometries to view galactic cosmic protons versus background components, the required integration time to reach a 3 sigma certainty in count rates is plotted in \cref{fig:int_times}. 

The coincidence mode here reduces integration times in the horns region and some of the SAA region significantly. Ideal count rates here are as low as possible, although as the full half-orbit as simulated was 59.0 minutes, meaning that the latitude and corresponding count rates in this half-orbit could vary significantly in just several minutes. Therefore a dashed line showing 1 minute in integration time is also plotted to show a rough limit for how low integration times might have be.

The integration times were calculated using \cref{eqn:int_time} below.

\begin{equation}
\label{eqn:int_time}
    T_{3\sigma} = 3^2 \times \frac{F_{signal} + \sum{F_{background}}}{F_{signal}^2}
\end{equation}

where $T_{3\sigma}$ is the required instantaneous integration time in seconds to get an observed signal-to-noise ratio $\leq$ 3, $F_{signal}$ is the count rate in cts / second of the signal that is being observed, and $\sum{F_{background}}$ is the sum of all count rates that are present but which aren't signal. In the cases given in \cref{fig:int_times}, the background for the GCR signal was exclusively the trapped particle components, while the background in the GLE cases was the GCR signal plus the trapped particle count rates.

\Cref{fig:int_times} shows an expected trend that in the single radiator case it takes significantly longer to achieve 3 sigma confidence in the horns regions and in the south Atlantic anomaly for each of the interplanetary proton cases. The integration times in the horns region for GCR and GLE05 cases are significantly improved in the horns region by the coincidence mode, and the region of count rates obscured by the SAA is slightly reduced. However, integration times are slightly increased in the regions without trapped particles, due to the reduction in signal count rates by a factor of 3 due to the loss of geometric viewing area. This therefore shows that while both the single radiator and coincidence modes/configurations can produce valuable observations of interplanetary protons, the single radiator mode is better for use in regions without significant trapped particle fluxes, while coincidence is required for observations in regions with strong trapped particle fluxes.

\section{Conclusions}
\label{sec:Conclusion}

The particle components experienced by a space-based Cherenkov detector are complex and both location-dependent and time-dependent, making the assessment of particle background challenging, particularly in the case of SEP detection. The research described in this paper has found that while a simple cubic fused silica Cherenkov detector should be able to measure both cosmic ray and SEP spectra, the horns region and SAA region are likely to obscure signals if mitigation methods are not used. The simple coincidence method investigated in this paper is very effective at removing trapped particle background in these regions, however it is interesting that it is unable to remove the background due to high energy proton components in the SAA. Its also interesting that there is also a proton component in the SAA originating from protons below the Cherenkov threshold, and the impacts of this component deserve future investigation.

\section{Funding}
\label{sec:Funding}

This research was carried out in collaboration with the European
Space Agency (ESA), under contract 4000139760/22/NL/CRS/my. This
work was also supported by the UK Space Agency under contract
UKSAG22\_0031\_ETP2-024.

\section{Declaration of competing interest}
\label{Declaration}

The authors declare that they have no known competing financial interests or personal relationships that could have appeared to influence the work reported in this paper.

\section{Data Availability Statement}
\label{sec:DataAvailability}

The notebooks used to generate all the plots and numbers used in this paper have all been included in the SpaceCherenkovSimulator Github repository at https://github.com/ssc-maire/SpaceCherenkovSimulator \cite{SpaceCherenkovSimulator}. The notebooks themselves, as well as plots and data used in this paper, are in the rough\_example\_notebooks subdirectory of this repository. Note that the data itself that was generated through simulations are stored in cache files within the directory structure; you may need some practice of using pickling within Python to access these. The simulations performed in the notebooks require an installed copy of Magnetocosmics\cite{MAGNETOCOSMICS} and GRAS\cite{santin2005gras} to run. The notebooks themselves require some import modifications to run properly, as when they were run to produce the plots in this paper, they used a local version of SpaceCherenkovRadiator, rather than the installeable package. SpaceCherenkovSimulator is the Python wrapper for GRAS for performing simulations of Cherenkov detectors as developed for this work. Others are welcome to use this package as part of their own work, or to fork it and add their own features, however note that as of writing the package has only been tested on one machine. If you do use it, please reference this paper and the package repository.



\bibliographystyle{elsarticle-num} 
\bibliography{bibliography}






\end{document}